\documentclass[]{spie}  

\usepackage{amsmath,amsfonts,amssymb}
\usepackage{graphicx}
\usepackage[colorlinks=true, allcolors=blue]{hyperref}

\title{Automated scheduler for the SOXS instrument: design and performance}

\author[a,b]{Laura Asquini}
\author[a]{Marco Landoni}
\author[d]{Dave Young}
\author[e]{Laurent Marty}

\author[s]{Stephen J. Smartt}

\author[a]{Sergio~Campana}
\author[f]{Riccardo~Claudi}
\author[e]{Pietro~Schipani}
\author[l]{Jani~Achrén}
\author[a]{Matteo~Aliverti}
\author[p]{José A.~Araiza-Durán}
\author[o]{Iair~Arcavi}
\author[f]{Federico Battaini}
\author[f]{Andrea~Baruffolo}
\author[g]{Sagi~Ben-Ami}
\author[a]{Andrea Bianco}
\author[g]{Alex Bichkovsky}
\author[p]{Anna	Brucalassi}
\author[g]{Rachel	Bruch}
\author[e]{Giulio~Capasso}
\author[f]{Enrico~Cappellaro}
\author[e]{Mirko~Colapietro}
\author[k]{Rosario~Cosentino}
\author[i]{Francesco~D'Alessio}
\author[a]{Paolo~D'Avanzo}
\author[e]{Massimo~Della~Valle}
\author[e]{Sergio	D'Orsi}
\author[k]{Rosario	Di Benedetto}
\author[f]{Simone	Di Filippo}
\author[g]{Avishay~Gal-Yam}
\author[a]{Matteo~Genoni}
\author[q]{Marcos~Hernandez}
\author[g]{Ofir	Hershko}
\author[m]{Jari~Kotilainen}
\author[m]{Hanindyo~Kuncarayakti}
\author[i]{Gianluca~Li~Causi}
\author[m]{Seppo~Mattila}
\author[k]{Matteo~Munari}
\author[a]{Giorgio Pariani}
\author[q]{Hector	Pérez Ventura}
\author[n]{Giuliano~Pignata}
\author[f]{Kalyan~Radhakrishnan}
\author[g]{Michael~Rappaport}
\author[f]{Davide~Ricci}
\author[a]{Marco~Riva}
\author[g]{Adam~Rubin}
\author[f]{Bernardo~Salasnich}
\author[e]{Salvatore	Savarese}
\author[r]{Maximilian	Stritzinger}
\author[k]{Salvatore~Scuderi}
\author[i]{Fabrizio~Vitali}
\author[f]{Ricardo~Zanmar~Sanchez}

\affil[a]{INAF – Osservatorio Astronomico di Brera-Merate, via E. Bianchi 46, I-23807 Merate (LC), Italy;}
\affil[b]{Dipartimento di Scienza e Alta Tecnologia, Università dell’Insubria, via Valleggio 11, I-22100 Como, Italy}
\affil[c]{INAF - Osservatorio Astronomico di Cagliari. Via della Scienza 5, Selargius (CA) - Italy}
\affil[d]{Astrophysics Research Centre, School of Mathematics and Physics, Queen's University Belfast, Belfast BT7 1NN, UK}
\affil[e]{INAF - Osservatorio Astronomico di Capodimonte, Salita Moiariello 16, Naples- Italy }
\affil[f]{INAF -- Osservatorio Astronomico di Padova, Vicolo dell’Osservatorio 5, I-35122, Padua, Italy }
\affil[g]{Weizmann Institute of Science, Herzl St 234, Rehovot, 7610001, Israel }
\affil[h]{Max-Planck-Institut für Extraterrestrische Physik, Giessenbachstr. 1, D-85748 Garching, Germany }
\affil[k]{INAF -- Osservatorio Astrofisico di Catania, Via S. Sofia 78 30, I-95123 Catania, Italy }
\affil[i]{INAF -- Osservatorio Astronomico di Roma, Via Frascati 33, I-00078 M. Porzio Catone, Italy }
\affil[l]{Incident Angle Oy, Capsiankatu 4 A 29, FI-20320 Turku, Finland }
\affil[m]{Finnish Centre for Astronomy with ESO (FINCA), FI-20014 University of Turku, Finland}
\affil[n]{Instituto de Alta Investigación, Universidad de Tarapacá, Arica, Casilla 7D, Chile}
\affil[o]{Tel Aviv University, Department of Astrophysics, 69978 Tel Aviv, Israel }
\affil[p]{INAF - Osservatorio Astrofisico di Arcetri.Largo Enrico Fermi 5, 50125 Florence - ITALY }
\affil[q]{FGG-INAF, TNG, Rambla J.A. Fernández Pérez 7, E-38712 Breña Baja (TF), Spain }
\affil[r]{Aarhus University, Ny Munkegade 120, D-8000 Aarhus, Denmark }
\affil[s]{University of Oxford, Keble Road, Oxford OX1 3RH}

\authorinfo{laura.asquini@inaf.it}

\begin{document} 
\maketitle

\begin{abstract}
We present the advancements in the development of the scheduler for the Son Of X-shooter (SOXS, \citenum{schipani2016new, schipani2018soxs}) instrument at the ESO-NTT 3.58-m telescope in La Silla, Chile. SOXS is designed as a single-object spectroscopic facility and features a high-efficiency
spectrograph with two arms covering the spectral range of 350-2000 nm and a mean resolving power of approximately R=4500. Its primary purpose is to conduct UV-visible and near-infrared follow-up observations of astrophysical transients, drawing from a broad pool of targets accessible through the streaming services of wide-field telescopes, both current and future, as well as high-energy satellites. The instrument is set to cater to various scientific objectives within the astrophysical community, each entailing specific requirements for observation planning, a challenge that the observing scheduler must address. A notable feature
of SOXS is that it will operate at the European Southern Observatory (ESO) in La Silla, without the presence of astronomers on the mountain. This poses a unique challenge for the scheduling process, demanding a fully automated algorithm that is autonomously interacting with the appropriate databases and the La Silla Weather API, and is capable of presenting the operator not only with an ordered list of optimal targets (in terms of observing constraints) but also with optimal backups in the event of changing weather conditions. This requirement imposes the necessity for a scheduler with rapid-response capabilities without compromising the optimization process, ensuring the high quality of observations and best use of the time at the telescope. We thus developed a new highly available and scalable architecture, implementing API Restful applications like Docker Containers, API Gateway, and Python-based Flask frameworks. We provide an overview of the current state of the scheduler, which is now ready for the approaching on-site testing during Commissioning phase, along with insights into its web interface and preliminary performance tests.
\end{abstract}


\section{INTRODUCTION}
\label{sec:intro}  

The Son Of X-Shooter (SOXS, \citenum{schipani2018soxs}) instrument is a single-object spectrograph with a spectral resolution of approximately R=4,500, designed to observe the southern sky from the New Technology Telescope (NTT) in La Silla, Chile. It will operate within the ESO La Silla-Paranal Observatory environment without an astronomer physically present on the mountain. Consequently, it necessitates the development of a web-based software capable of autonomously organizing and managing nightly observations, both in advance (for scientists' approval and scrutiny) and in real-time. These processes must adhere to and optimize observational constraints, while also periodically checking current weather conditions on the mountain. Additionally, the scheduler must integrate targets from both the SOXS Consortium and regular ESO proposals, respecting the 50/50 Guaranteed Time of Observation agreement. The Targets of Opportunity (ToOs) from the SOXS Consortium will be managed through the Marshall application\cite{smartt2015pessto}, while other ESO users will use the Phase 2 (P2) web page. Throughout and at the end of each night, the scheduler needs to track the instrument's operations, including the percentage of completed observations, hit/miss rate of transient events, and general observing conditions for each target. It will automatically communicate with the official ESO Phase 2 tools, creating and verifying Observing Blocks (OBs) from the targets in the standard ESO format, and sending them to fill the Visitor Execution Sequence (VES) based on target visibility, priority, and current weather conditions accessed through the Astronomical Site Monitor (ASM) Weather API. We refer the reader to previous works [\citenum{asquini2022dynamic, landoni2020soxs}] for the first presentation of the software and a thorough description of its components. In this paper, we will focus on the performance of the real-time night managing algorithm, which is the crucial part of the nightly operations. The paper is organized as follows: in Section \ref{sec:2} we will briefly summarize the main features of the software. In Section \ref{sec:3} we will describe the night-managing algorithm, while in Section \ref{sec:4} we will focus on its performance. We draw our conclusions in Section \ref{sec:5}.

\section{Main Features}
\label{sec:2}
In this Section we summarize the building blocks of the software, that have detailed description in previous works [\citenum{asquini2022dynamic, landoni2020soxs}].

\begin{figure} [ht]
   \begin{center}
   \includegraphics[width=0.7\textwidth]{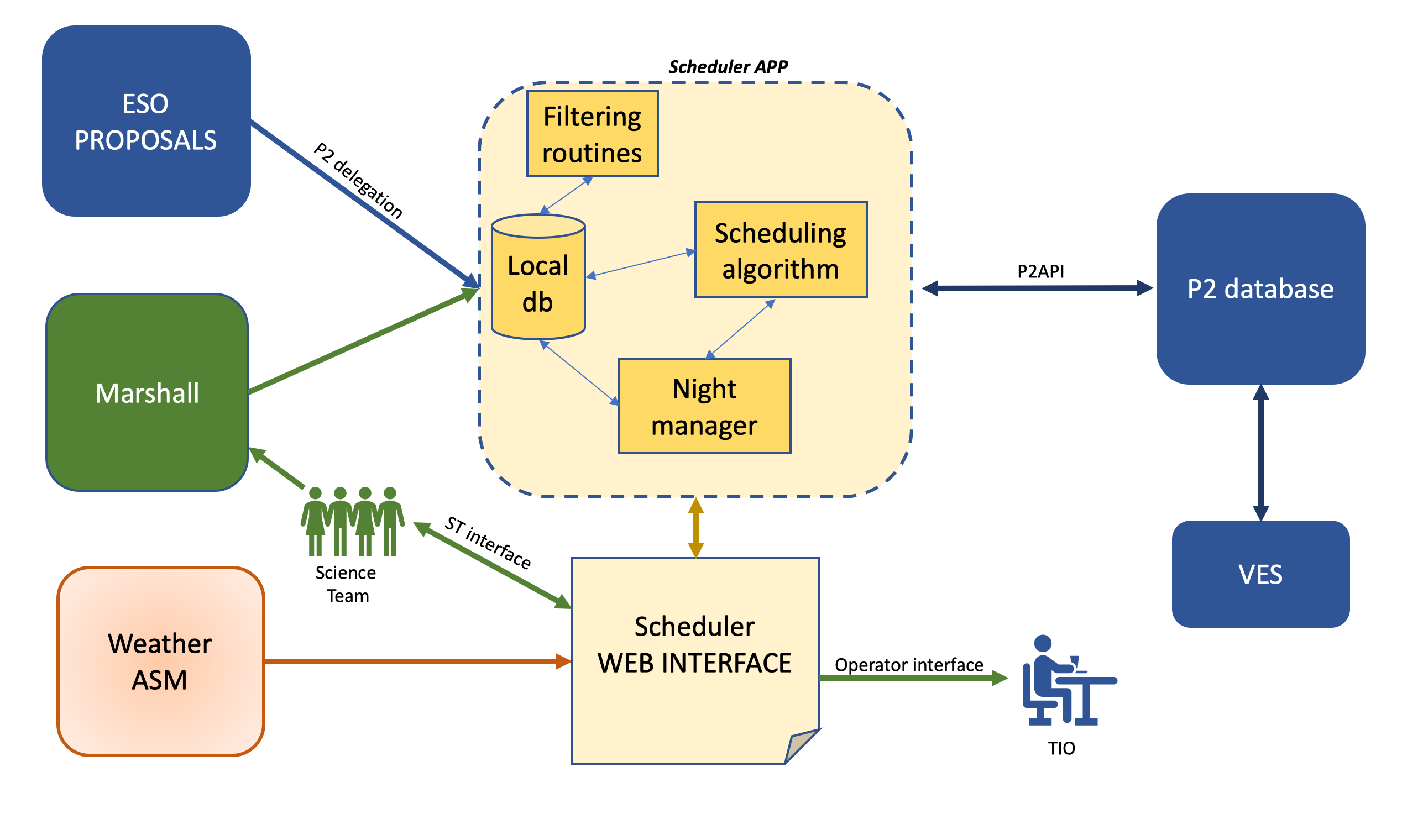}
   \end{center}
   \caption{The overall workflow of the SOXS scheduler.}

   \label{fig:sched} 
   \end{figure}

The scheduler is implemented as a RESTful API using the Flask micro-framework, serving as an intermediary between the Marshall and the ESO P2 system. OBs are created via the SOXS scheduler and stored in a MySQL database. The core code is adaptable for any ground-based instrument, requiring only two additional files for instrument and telescope specifics.
The scheduling algorithm is written in Python 3, utilizing Astropy\cite{price2018astropy} libraries, particularly Astroplan, to compute observational constraints and employ a priority scheduler. The algorithm's workflow is divided into two steps: the first processes the computational workload and selects observable targets for the night, using scripts for new and existing transients. The second step is the actual scheduling process.
The data stream starts with targets fed into the SOXS API by the Marshall or ESO P2, which are then processed daily to ensure up-to-date information. The scheduler creates an optimized schedule, which is sent to the P2 API for validation and then to the VES. The system continuously requests feedback on observation status from P2 and checks weather conditions via the ASM API. If issues arise, the automatic night manager adjusts the schedule or selects alternative OBs.
Figure \ref{fig:sched} shows the overall structure of the scheduler and its data flow.

\subsection{Filtering}
Target filtering is essential to reduce the workload on the scheduler. Two scripts were developed for this purpose: one processes new OBs, and the other updates existing ones. These scripts operate continuously, day and night, periodically checking the database for their respective OBs.

The algorithm, detailed in a previous paper, represents the night using 1D arrays with five-minute time slots. Each target has an associated observability array, where each slot is marked as zero (not observable) or one (observable) based on the convolution of observing constraints. Additionally, each target is graded on its likelihood of a good observation for the night, using a system adapted from ESO's observation planning.

The processing script evaluates each target for five consecutive nights, while the updating script discards the past night and adds the fifth night from the current day. This ensures a wide availability of targets for both the scheduler and the scientific team at all times.

\subsection{Scheduling}
The scheduler works on three types of OB: ESO OBs (high-priority, coming from the general ESO public), follow-up OBs (high-priority, coming from the SOXS Consortium), and classification OBs (low-priority, targets that are unclassified and are public, unless a scientist from the Consortium claims them as follow-up). 
To manage the scientific prioritization of targets, particularly for follow-up and ESO proposals, the built-in Astroplan\cite{morris2018astroplan} priority scheduler is used as the first step to handle top-priority targets. This scheduler optimizes observing times for each object, evaluating constraints with a fine one-minute time resolution. While efficient, this process is computationally intensive, so it is reserved for high-priority targets.

The remaining time is managed with an algorithm called the Graded Gap Filler (GGF\cite{asquini2022dynamic}) using 1D boolean arrays representing the night in five-minute slots. The scheduler fills gaps by finding the first free slot and matching it with suitable observable targets from a pre-sorted list, based on their grades and required slots. The algorithm selects the best candidate for each slot and updates the night array accordingly, continuing until the night is full or no suitable targets remain.

During an observing night, the scientific team can tweak the schedule through a web interface but does not need direct access to the algorithms. At the beginning of the night, or when the team approves a schedule, a new night is opened via the interface. The scheduler proposes an optimized schedule, respecting scientific priorities and observation-quality constraints. The team can swap OBs with alternatives provided by the scheduler or manually add targets from the ``observable obs'' page. Once the schedule is finalized and approved, it synchronizes with P2 for verification, and the first OB is sent to the Visitor Execution Sequence (VES), which is the list of ordered OB that the observatory system checks for the night schedule.

   \begin{figure} [ht]
   \begin{center}
   \includegraphics[width=0.7\textwidth]{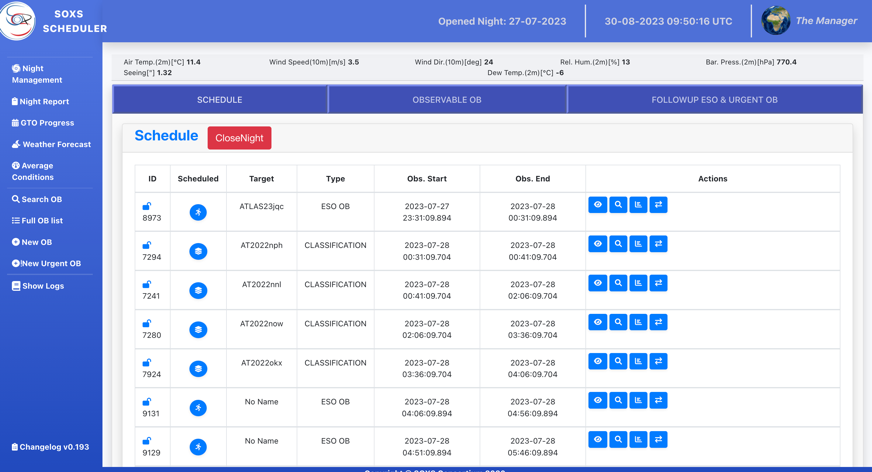}
   \end{center}
   \caption{The interface for the Science Team. It features various tools to aid the SOXS scientists to tweak and perfect the schedule.}

   \label{fig:scif} 
   \end{figure} 

\section{Night Managment}
\label{sec:3}

In this project, a key challenge for the scheduler is to autonomously monitor and respond to real-time events, such as weather changes during an observing night. This task is managed by the ``Next OB'' endpoint, operated by the telescope instrument operator (TIO). The TIO requests a new OB from the scheduler after each one is completed or fails. Upon receiving a request, the scheduler checks current weather conditions via a weather API and verifies time constraints before approving the next OB for observation and sending it to the VES.

Three main event categories that might trigger schedule adjustments have been identified:
\begin{itemize}
    \item Delays in executing observations;
    \item Previously failed OBs;
    \item Weather changes
\end{itemize}

To address these scenarios, three distinct ``recovery loops'' have been implemented. Each recovery loop is an algorithm designed to handle specific issues during the observing process. These loops are central to the ``Next OB'' endpoint's functionality. The approach aims to adhere closely to the initially approved schedule, minimizing conflicts with the time share between users and the Science Team’s decisions.

\begin{figure} [ht]
   \begin{center}
   \includegraphics[width=0.7\textwidth]{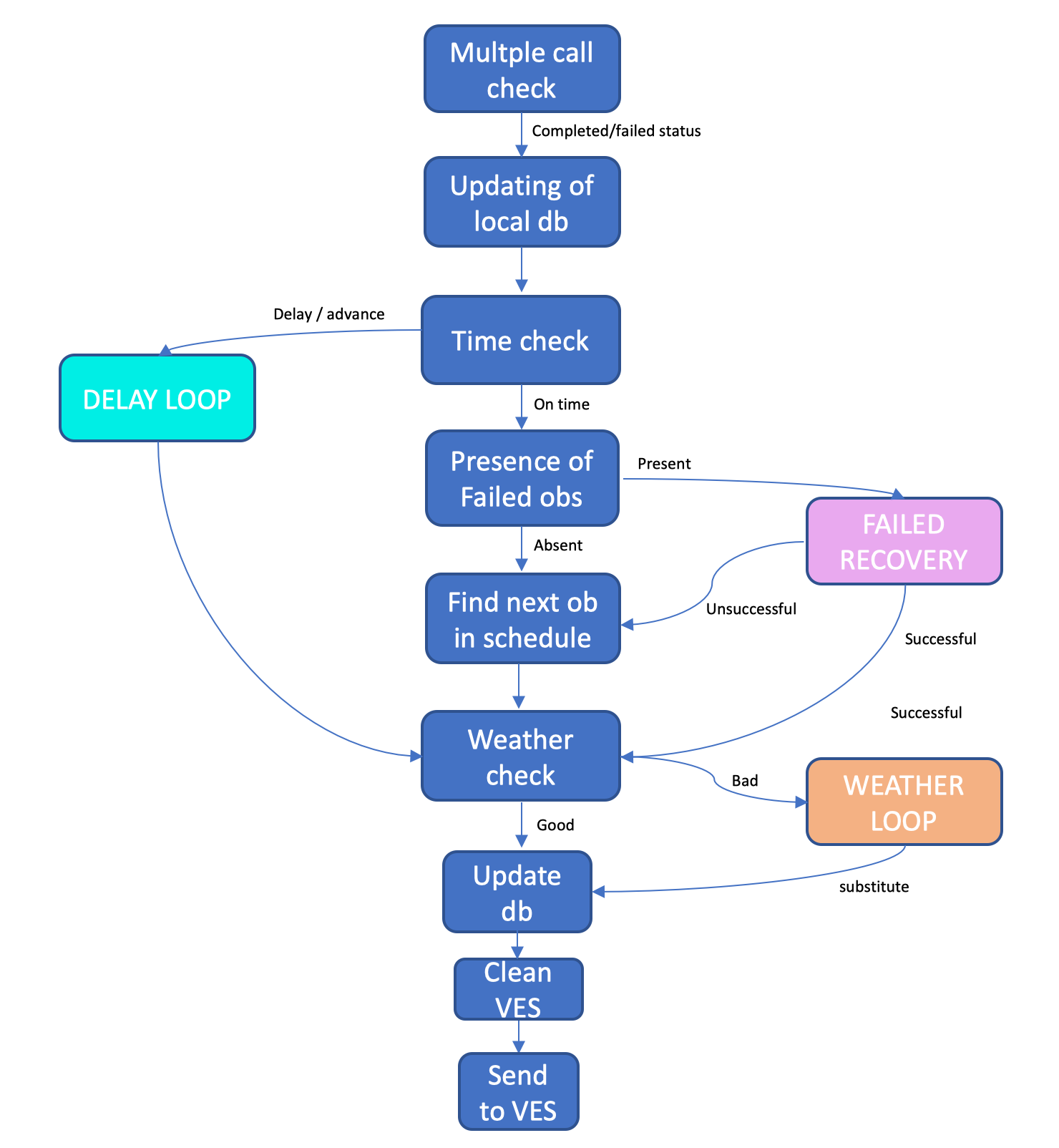}
   \end{center}
   \caption{The workflow of the ``Next OB'' algorithm.}

   \label{fig:nextob} 
   \end{figure}

The process for handling OBs in the VES involves several checks and updates, which are shown in Figure \ref{fig:nextob}:
\begin{itemize}
    \item OB Status Check: The first check determines if the current OB in the VES is marked as completed (status X) or failed (status A) in the P2 database. If the OB is still in progress (status ``+'' for ``Accepted'' or ``S'' for ``in observation''), it is returned without further checks to avoid multiple-call errors.
    \item Status Update: The OB status is updated in the local database.
    \item Time Check: A check is performed to ensure there are no major delays or advances. If there are, the Delay Loop is triggered. It will reschedule the remaining OBs, prioritizing the observation of high-priority targets, and filling any gaps left.
    \item Failed OB Check: If the schedule is on time, the presence of failed OBs is checked. If there are failed OBs, the Failed OB Recovery Loop is triggered. It will try and rearrange the remaining time, inserting the previously failed OBs while still maintaining the high-priority ones.
    \item Subsequent OB Check: If no OBs have failed, the next OB is identified, and its status is checked to ensure it is the correct choice.
    \item Weather Check: Current weather conditions are retrieved via the API and checked against the OB requirements. If requirements are not met, the Weather Check Loop is triggered. It will find a suitable OB for the current weather conditions, to fill the gap left by the current candidate OB. 
    \item Failed OB Management: if a priority target OB fails, it is added to the ``failed obs'' table, so that it will be checked again later for recovery. Classification targets, which are interchangeable, are not added to this list. If the schedule has undergone any recovery loops, the new schedule is saved, and the previous one is archived in the ``history schedule'' table.
    \item VES Update: If everything is in order, the VES is cleaned, and the current OB is sent to the VES for observation.
\end{itemize}

\section{Performance}
\label{sec:4}
To evaluate the overall quality of the resulting schedules, a three-month simulation was conducted. This simulation incorporated artificial time progression, simulated weather conditions, and introduced a thousand new targets each night using recorded transient alerts from the ePESSTO collaboration\cite{smartt2015pessto} in 2018. Targets were generated with randomized magnitudes (not exceeding the maximum allowed magnitude of 21, valid for SOXS) to handle 1000 potentially observable new targets per night. Follow-up requests from the SOXS Consortium and ESO users shared 50\% of telescope time, averaging 10 requests per day.

Randomized weather conditions were based on ESO data for Cerro Paranal in the Atacama desert. Unforeseen events, such as telescope malfunctions or unregistered bad weather, were simulated as ``delays'' occurring approximately every 20 OBs (1-2 days), ranging from 30 minutes to two hours. The simulation automated scheduler functionalities, including target creation, database cleanup, filtering processes, schedule creation, and night progression. The stress-test was conducted under worst-case conditions, such as during Chilean summer with shorter nights, starting January 7, 2018.

The simulation showed that, even under undesirable conditions (overbooking, weather changes, frequent delays), the scheduler managed to optimize 54.4\% of requested targets, observing them within a day of creation. An additional 18.2\% of OBs were recovered during night rearrangements and observed within 30 days, while 27.4\% could not be recovered. These results highlight the importance of scientific prioritization by the Science Team to ensure high-priority OBs are observed on the scheduled night.

Regarding observation quality, the most significant metric was "delta airmass." Approximately 90-95\% of targets were observed within a 0.2 airmass of their optimal position. The first schedule had a mean delta airmass of 0.1$\pm$0.2, which stabilized at 0.1$\pm$0.1 after night management, maintaining consistency even with a heavier load on the GGF compared to the Priority scheduler. The average response time for the process, from clicking the next OB button to viewing it on the operator interface, was less than 3 seconds with a typical fiber optic internet connection.

\section{Conclusions}
\label{sec:5}
The development and implementation of the scheduler for the SOXS system has demonstrated significant robustness and efficiency, even under challenging conditions.
The scheduler successfully handles real-time events such as weather changes through the ``Next OB'' endpoint, ensuring smooth operation throughout the observing night. This capability is critical for maintaining the integrity of the observation schedule despite unforeseen events. Specificallt, the recovery loops (Rescheduling, Failed OB Recovery, and Weather Check) effectively address delays and failed observations, maintaining a high success rate for target observation.
The three-month simulation incorporated real-time weather data and a high volume of targets, and the scheduler demonstrated its capability to manage heavy loads and adverse conditions. Despite overbooking and frequent delays, the scheduler optimized 54.4\% of the targets within a day and recovered an additional 18.2\% within 30 days. The resulting observations consistently proved to be of high-quality, with 90-95\% of targets observed within a 0.2 airmass of their optimal position. Finally, the average response time of the system was less than 3 seconds under normal internet conditions. This quick rectivity is crucial for real-time observation adjustments and maintaining a seamless workflow.

\bibliography{report} 

\begin{thebibliography}{1}

\bibitem{schipani2016new}
Schipani, P., Claudi, R., Campana, S., Baruffolo, A., Basa, S., Basso, S.,
  Cappellaro, E., Cascone, E., Cosentino, R., D'Alessio, F., et~al., ``The new
  soxs instrument for the eso ntt,'' in [{\em Ground-based and Airborne
  Instrumentation for Astronomy VI}{\nolinebreak\hspace{0.1em}]},   {\bf 9908},
   1268--1277, SPIE (2016).

\bibitem{schipani2018soxs}
Schipani, P., Campana, S., Claudi, R., K{\"a}ufl, H., Accardo, M., Aliverti,
  M., Baruffolo, A., Ami, S.~B., Biondi, F., Brucalassi, A., et~al., ``Soxs: a
  wide band spectrograph to follow up transients,'' in [{\em Ground-based and
  Airborne Instrumentation for Astronomy VII}{\nolinebreak\hspace{0.1em}]},
  {\bf 10702},  96--107, SPIE (2018).

\bibitem{smartt2015pessto}
Smartt, S., Valenti, S., Fraser, M., Inserra, C., Young, D., Sullivan, M.,
  Pastorello, A., Benetti, S., Gal-Yam, A., Knapic, C., et~al., ``Pessto:
  survey description and products from the first data release by the public eso
  spectroscopic survey of transient objects,'' {\em Astronomy \&
  Astrophysics}~{\bf 579},  A40 (2015).

\bibitem{asquini2022dynamic}
Asquini, L., Landoni, M., Young, D., et~al., ``Dynamic scheduling for soxs
  instrument: environment, algorithms and development,'' in [{\em Software and
  Cyberinfrastructure for Astronomy VII}{\nolinebreak\hspace{0.1em}]},   {\bf
  12189},  106 (2022).

\bibitem{landoni2020soxs}
Landoni, M., Young, D., Marty, L., Asquini, L., Smartt, S.~J., Trombetta, A.,
  Campana, S., Claudi, R., Schipani, P., Aliverti, M., et~al., ``The soxs
  scheduler for remote operation at lasilla: concept and design,'' {\em arXiv
  preprint arXiv:2012.12677}  (2020).

\bibitem{price2018astropy}
Price-Whelan, A.~M., Sip{\H{o}}cz, B., G{\"u}nther, H., Lim, P., Crawford, S.,
  Conseil, S., Shupe, D., Craig, M., Dencheva, N., Ginsburg, A., et~al., ``The
  astropy project: building an open-science project and status of the v2. 0
  core package,'' {\em The Astronomical Journal}~{\bf 156}(3),  123 (2018).

\bibitem{morris2018astroplan}
Morris, B.~M., Tollerud, E., Sip{\H{o}}cz, B., Deil, C., Douglas, S.~T.,
  Medina, J.~B., Vyhmeister, K., Smith, T.~R., Littlefair, S., Price-Whelan,
  A.~M., et~al., ``Astroplan: an open source observation planning package in
  python,'' {\em The Astronomical Journal}~{\bf 155}(3),  128 (2018).

\end{thebibliography}
\bibliographystyle{spiebib} 

\end{document}